\documentclass[journal]{IEEEtran}
\ifCLASSINFOpdf
  \usepackage[pdftex]{graphicx}
\else
\fi
\usepackage{amsmath}
\usepackage{import}
\usepackage{amsfonts}
\usepackage{cite}
\usepackage{color}

\newcommand{\refeq}[1]{Eq.\,(\ref{#1})} 
\begin{document}
%
\title{Deep Learning Based Computed Tomography\\
Whys and Wherefores}
%
%
%

\author{Shabab Bazrafkan,
        Vincent Van Nieuwenhove,
        Joris Soons,
        Jan De Beenhouwer,
        and Jan Sijbers 
\thanks{S. Bazrafkan, J. De Beenhouwer and J. Sijbers are with imec Visionlab, Department of Physics, University of Antwerp,
Antwerp, Belgium e-mail: \{shabab.bazrafkan\},\{jan.debeenhouwer\},\{jan.sijbers\}@uantwerpen.be.}
\thanks{V. Van Nieuwenhove and J. Soons are with Agfa NV, Mortsel, Belgium
email: \{vincent.vannieuwenhove\},\{joris.soons\}@agfa.com.}}
\maketitle



%
\IEEEpeerreviewmaketitle

\section{How Deep Learning Became Deep Learning}
%
%
%
%
\IEEEPARstart{I}{n} the last few years with the emerging of affordable parallel processing hardware and free and open source frameworks, a new type of machine learning approach named "Deep Learning" drew an extensive amount of attention. The wave originally started with a Deep Neural Network named AlexNet \cite{ALEXNET} in 2012 which won the ImageNet Large-Scale Visual Recognition Challenge (ILSVRC). This new method achieved an incredible 8.8\% less top five test error compared to the second best method. After 2012, the DNN models were the usual winners of every year's ILSVRC. Several amazing architectures were introduced for object detection since 2012 including but not limited to ZF Net \cite{ZFNET}, VGG Net \cite{VGG}, GoogleNet \cite{INCEPTIONRESNETV2}, and Microsoft's ResNet \cite{RESNET} that won the ILSVRC in 2015 with an error rate of 3.6\% which is almost as twice as better than the human accuracy. \\
 
While the deep learning science started with object classification, the wave did not stop just there. Object detection was another big application wherein DNNs took a large step forward. Works such as Region Based CNNs (RCNN) \cite{RCNN}, Fast RCNN \cite{FASTRCNN}, Faster RCNN \cite{FASTERRCNN}, You Only Look Once (YOLO) \cite{YOLO},  YOLO9000 and YOLOv2 \cite{YOLO9000} and Mask Region-based Convolutional Network (Mask R-CNN) \cite{MASKRCNN} are methods which were designed to provide a bounding box around an object in their input image and also classify the object.\\
 
While these works mix a regression problem (bounding box location) with a classification solution (object recognition), there are applications on multidimensional classification use cases such as semantic segmentation applications. For example, the celebrated SegNet model is designed to map its input image to an output of the same size in which the output pixels correspond to one of 11 classes. This is while the networks presented in \cite{MYIRIS,MYIRIS2} are trained to perform binary classification for segmenting low-quality iris images. Considering medical applications, the U-net model \cite{UNET} is trained to perform binary segmentation of neuronal structures which is also utilized to perform segmentation in transmitted light microscopy images (phase contrast and DIC).\\
 
Classification and Regression are not the only problems for which DL provides a solution. In 2014, DNNs has been utilized in estimation theory by the introduction of Generative Adversarial Networks (GAN) \cite{GAN}. GANs are successful implementations of deep generative models which learn the data distribution and draw random samples from the learned distribution. There are multiple variations such as WGAN \cite{WGAN}, EBGAN \cite{EBGAN}, BEGAN \cite{BEGAN}, VAR+GAN \cite{VARGAN} and VAC+GAN \cite{VACGAN}, which have evolved from the original GAN by adjusting the loss function and/or the network architecture.\\

Alongside all different applications of DNNs, their impact on the medical imaging sciences is substantial. This article is mainly around the impact of DNNs on the CT imaging in general and low dose CT reconstruction in particular since recently DNNs have been widely used in low dose CT use cases. The technologies are getting so progressive that recently the first market stage deep learning based low dose CT reconstruction \cite{AICE} have been introduced. This indicates the importance of data-driven methods such as Deep Learning even in sensitive markets such as medical imaging. The success of these methods in providing valuable input to radiologists changes the path where the imaging technology heads. In the Next section a brief explanation of CT imaging is described followed by a discussion of the influence of DL in the CT imaging in section \ref{sec:DNNinCT}. In section\ref{sec:future} a more detailed explanation of the future of CT imaging is described and the last section includes the conclusion and the discussion. 


\section{CT Imaging}
Computed tomography (CT)  is well-known imaging technique that allows for non-invasive visualization of the interior of an object.  It is widely used in many applications such as medical imaging \cite{Bach2012,Kubo08}, non-destructive testing \cite{DeSchryver2016}, industrial metrology \cite{HILLER2016}, food industry \cite{Schoeman2016,Janssens2015}, and security \cite{SHIKHALIEV2018}.
In CT,  X-ray radiation is used to acquire a number
of two dimensional (2D) images of an object from many different view points. From these images, using reconstruction software, a three dimensional (3D) model of the object’s 
internal structure is computed and subsequently analyzed.
In this section, we will shortly describe the principle of X-ray CT imaging and image reconstruction. 

\subsection{X-rays: matter interaction and detection}
\label{sec:introduction:acquisitionProcess:Xray}



When an X-ray beam passes through an object, its intensity decreases due to physical mechanisms such as the photo-electric effect or elastic or inelastic scattering. 
Let $I_0$ denote the intensity of a monochromatic X-ray beam that leaves the X-ray source. 
Then the intensity of the X-ray beam at position $s$ on the detector after passing through the object along a line $L$ oriented at angle $\theta$  is given by:
\begin{equation}
I_\theta(s) = I(0) e^{-\int_L \mu(\eta)d\eta} . \label{eq:1:derBeerLambert10}
\end{equation}
\refeq{eq:1:derBeerLambert10} describes the relationship between the observed intensity at the detector side and the unknown attenuation coefficients $\mu$ the X-ray beam passed through. Log-normalization of this detected intensity yields the projection value
\begin{equation}
p_\theta(s) = -\ln\left(\frac{I_\theta(s)}{I_0}\right) =
\int_L \mu(\eta)d\eta .
\label{eq:1:BeerLambertLinearForm}
\end{equation} 
which linearly relates to the (unknown) attenuation coefficients of the object. 

The main purpose of CT reconstruction methods is to recover the object's attenuation coefficients $\mu(.)$ from given projection data $p(.)$. In what follows, we describe commonly used reconstruction methods to recover the object's attenuation values from  projection data $p_\theta(s)$ measured by directing the X-ray beam at different angles $\theta$ through the object. 


\subsection{Analytical reconstruction methods} \label{sec:introduction:reconstructionMethods:analytical}
In the analytical approach, the object's attenuation distribution is described as a function $f: \mathbb{R} \times \mathbb{R} \rightarrow \mathbb{R}$ that maps the spatial coordinate $(x,y)$ to its corresponding local attenuation coefficient $\mu$. 

\subsubsection{Filtered Back Projection (FBP)}
\label{sec:introduction:reconstructionMethods:analytical:FBP}
The Filtered Back Projection (FBP) reconstruction method is based on the following analytical formula:
\begin{equation}
f(x,y) = \int_0^{\pi}\left\{ \int_{-\infty}^{\infty}P_{\theta}(q)|q|e^{2\pi i q (x \cos\theta + y\sin\theta)}dq \right\} d\theta .
\label{eq:1:FBPformula}
\end{equation}
where $P_{\theta}(q)$ denotes the Fourier transform of the projection data $p_{\theta}(s)$. 
As can be observed from \ref{eq:1:FBPformula}, the FBP formula gives rise to a simple two step approach for calculating a reconstruction of the scanned object based on the measured projection data \cite{VanEyndhoven2018}:\\

\begin{enumerate}
	\item Filter the projection data $p_{\theta}(r)$ by multiplying its Fourier transform $P_{\theta}(q)$ with $|q|$ and calculating the inverse Fourier transform. This step corresponds to the inner integral in \ref{eq:1:FBPformula}.\\
	
	\item For a particular location in the image domain $(x,y)$, sum up all the filtered projection data that corresponds to the lines $x \cos\theta + y\sin\theta$ with $\theta \in [0,\pi]$. This step corresponds to the outer integral in \ref{eq:1:FBPformula}.\\
\end{enumerate} 

The FBP representation assumes that projection data is available from all angles. If indeed many X-ray projections from all angles are available, FBP generally leads to high quality reconstructions.  If these two conditions are not satisfied (e.g. in case of limited angle scanning or related missing data problems), severe streaking artefacts appear in the reconstructed image.

\subsection{Algebraic reconstruction methods} 
\label{sec:introduction:reconstructionMethods:algebraic}
\begin{figure}[h!]
	\centering
	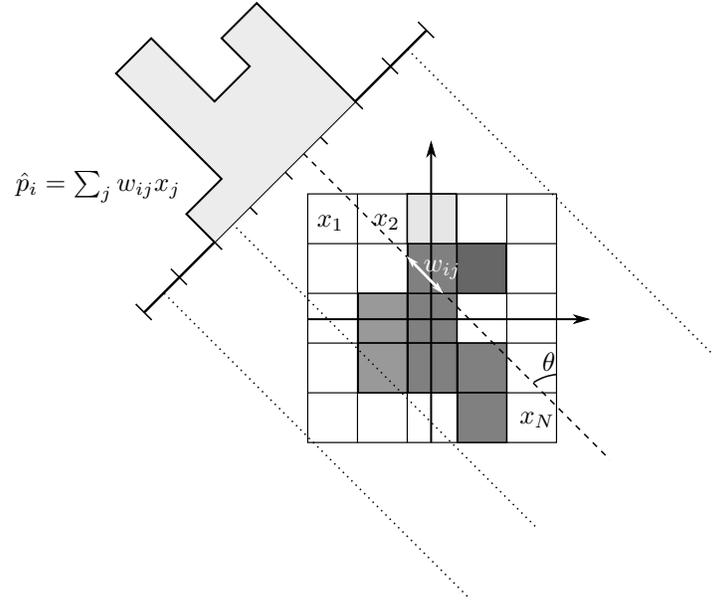
	\caption{The Discrete Projection Model}
	\label{fig:1:ProjModelDiscrete}
\end{figure}

A class of reconstruction methods that are better suited to cope with deviations from the above conditions are algebraic reconstruction methods (ARMs). ARMs methods rely on a discrete model of the object $\{x_j\}$, as shown in Fig.\ref{fig:1:ProjModelDiscrete}. Their basic scheme is to iteratively minimize the difference between the computed forward projection of the discrete image $\{\hat p_i\}$ with the observed projection data $\{p_i\}$, where the object is updated based on the backprojected difference. Thereby,  $\hat p_i = \sum_j w_{ij}x_j$, with $w_{ij}$ denoting the contribution of object pixel $x_j$ to the detector pixel $p_i$.

A commonly used ARM is the SIRT algorithm, in which, in each iteration $k$, the current estimate of the image, $\{x_j^k\}$, is updated as follows: 
\begin{equation}
x_j^{(k+1)} = x_j^{(k)} + \frac{1}{\sum_{i} w_{ij}}\sum_{i}\left( \frac{w_{ij}(p_i-\sum_{h}w_{ih}x_h^{(k)})}{\sum_{h} w_{ih}}\right) 
\label{eq:1:SIRTpixelwise}
\end{equation}

An important advantage of ARMs is that, if prior knowledge is available such as shape, object support, material density, or sparseness in some transform domain, this information can be easily integrated in such an iterative reconstruction scheme. As a result the quality of the reconstructed image can substantially be improved compared to plain FBP. One of the most important drawbacks, however is their computational load and slow convergence, which is the main reason why they are not commonly used in industrial applications.

\section{Deep Learning in CT}
\label{sec:DNNinCT}
The most studied Deep Learning application in CT imaging is in the reconstruction pipeline for low dose CT use cases. The main reason is the difficulty in finding the exact artefact model caused by the limited angle projections. These artefacts although seem similar in different reconstructions are nontrivial to predict without knowing the exact geometry and material properties of the object to be scanned. This is a dead loop since the goal of the imaging is to measure such properties. It is worthwhile to mention the data-driven methods provide a powerful tool to learn any kind of pattern which occur in similar shapes and intensities repetitively throughout any data stream. DNNs are able to learn a set of solutions from the training set and generalize it to a set of data which they never observed before \cite{CEmag1}. In the case of the low dose CT imaging, they give promising results in removing the streaking artefacts from the reconstructed images.\\
 
In general, DNNs provide solutions in CT reconstruction in two main different ways. First, they produce a post-processing tool to remove the artefacts caused by low dose imaging. Second, they present an end to end model which translates the sinogram space into the image space. Both of these approaches are explained in the following sections.

\subsection{DNNs as helping hands}
\label{sec:helpinghand}
If you already came across a deep learning article in the low dose CT reconstruction, it is most probably a DNN, utilized as a post-processing step to remove the noise and artefact from other reconstruction techniques. In fact, considering the literature, the biggest misleading lyes in the title of the articles which give the impression that the DNN is performing the reconstruction task. This is while in almost all of these papers the deep learning technique is used as an auxiliary step after the actual reconstruction from classical methods such as FBP or SIRT. For example in \cite{pelt2018improving}, the authors exploit a Mixed-Scale Dense (MSD) Convolutional Neural Network \cite{MSDorig} to remove the artefacts from the FBP reconstruction. In \cite{chen2017low,chen2017low2} other fully convolutional models have been used to perform the exact same task. This is while other approaches such as \cite{kang2017wavelet,kang2018deep} perform this job in the wavelet and contourlet space using fully convolutional DNNs.\\

There is another approach known as NNFBP presented in \cite{pelt2013fast} where a neural network is exploited to learn a filter bank for the FBP method and a weighted sum of several FBP reconstructions are calculated as the final result for a given sinogram.\\
 
There are also some sophisticated approaches like the work presented in \cite{yang2018low} wherein a GAN objective (on Wasserstein distance) is imposed on the network that forces it to generate an output which follows high-quality CT images distribution. This is while the network is also obligated to reduce a perceptual loss between its output and the ground truth which this loss is produced by another pre-trained network known as VGG \cite{VGG}. Perceptual losses are designed to produce visually friendly results and do not impose any structural and/or pixel level correctness to the reconstructed image. This is an important issue which should be considered in a detailed level if these methods are getting any market attention specially in medical imaging where sensitive decisions are made based on the acquisitions.\\

\subsection{DNNs as an end to end solution}

It is very tempting to be able to provide an end to end solution in CT reconstruction using data driven methods. This means that the model learns the mapping between the sinogram signal and image space entirely based on training data. The AUTOMAP method originally presented in \cite{zhu2018image} provides such an end to end solution. Since there is no obvious one to one, pixel level correspondence from sinogram space to the image space signals, especially in the low dose scenarios, the AUTOMAP technique exploits two fully connected layers at the very beginning of the model. The fully connected layers give the opportunity to each sinogram sample to contribute to all the pixels in the image space. The first layer maps the input sinogram into a signal with the size of the output, and the second layer maps this tensor to another signal with the same size. These two layers contribute to almost all learnable parameters in the network. Let's continue the discussion with an example. \\

Consider a 2-dimensional case where the X-ray sensor consists of 512 pixels and 128 signals in different projection angles are taken from the objects. In this case, the sinogram space signal will be an image with the dimensions $128\times 512$ pixels. If the reconstructed image considered to be $512\times 512$ pixels then the first two layer of AUTOMAP model consist of $(128\times 512^3+512^4)$ which is around $8.5\times 10^{10}$ learnable parameters. Most of the current deep learning platforms and also GPU architectures correspond 32 bit to each variable. This means that in such a simple scenario the AUTOMAP model needs more than 340 Gigabytes of memory to store its variables. The other problem is that training of such a huge model needs an extremely large database with enough variations to avoid overfitting. It is worthwhile to note we are only considering 2-dimensional use cases. For the real life 3-dimensional scenarios the needed memory and computing power rise exponentially. The AUTOMAP model also consists of two convolutional layers and an output layer.\\

\begin{figure}[!t]
\centering
\includegraphics[width=3.5in]{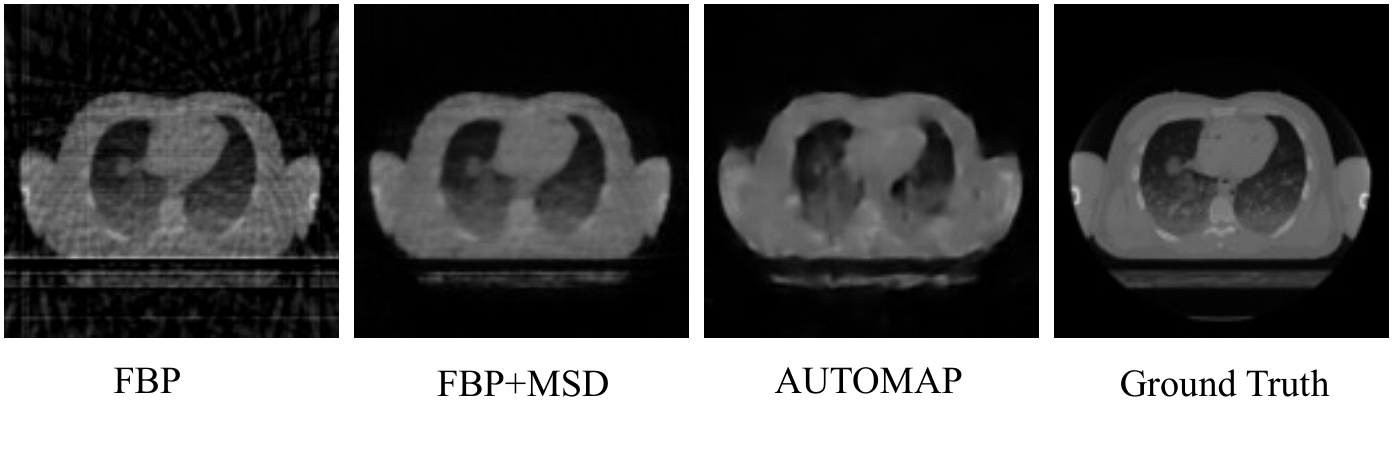}%
\caption{Reconstruction of low dose CT using FBP, repaired FBP, and AUTOMAP methods and the Ground Truth.}
\label{fig:results}
\end{figure}

In order to investigate this technique in the low dose scenarios, we trained an AUTOMAP model for a low dose CT problem with image size of $128\times 128$ and 20 projections with parallel beam geometry. The model was trained on more than 45000 images from National Cancer Institutes Clinical Proteomic Tumor Analysis Consortium Pancreatic Ductal Adenocarcinoma (CPTAC-PDA)\footnote{https://wiki.cancerimagingarchive.net/display/Public/CPTAC-PDA} database. We also trained an MSD network to reproduce the results of the method presented in \cite{pelt2018improving} which is explained in section \ref{sec:helpinghand}. The models are tested on samples from Visible Human Project CT Datasets\footnote{https://mri.radiology.uiowa.edu/visible human datasets.html}. The MSD network tends to remove the artefacts induced by the FBP algorithm in the reconstruction process. This observation is conducted to compare the end to end method AUTOMAP to a method where DNNs are used as an auxiliary step in the reconstruction pipeline for low dose CT. Figure \ref{fig:results} shows the reconstructions on a test sample for the original FBP method, the FBP output repaired using the trained MSD network, and AUTOMAP output alongside with the ground truth.\\

The AUTOMAP method is able to reconstruct the general shape of the object while the details are tangled into each other. The main reason is the fact that this approach totally ignores the geometry of the scanning process which plays a crucial role in FBP and consequently FBP+MSD methods. In ill-posed problems such as low dose CT reconstructions, any auxiliary information can make a difference in the final results. In the current case, the knowledge about scanning geometry induces auxiliary information which helps the model to retain more detailed structures in the final reconstruction.

\section{How the Future Looks Like?}
\label{sec:future}
\begin{figure*}[!t]
\centering
\includegraphics[width=7in]{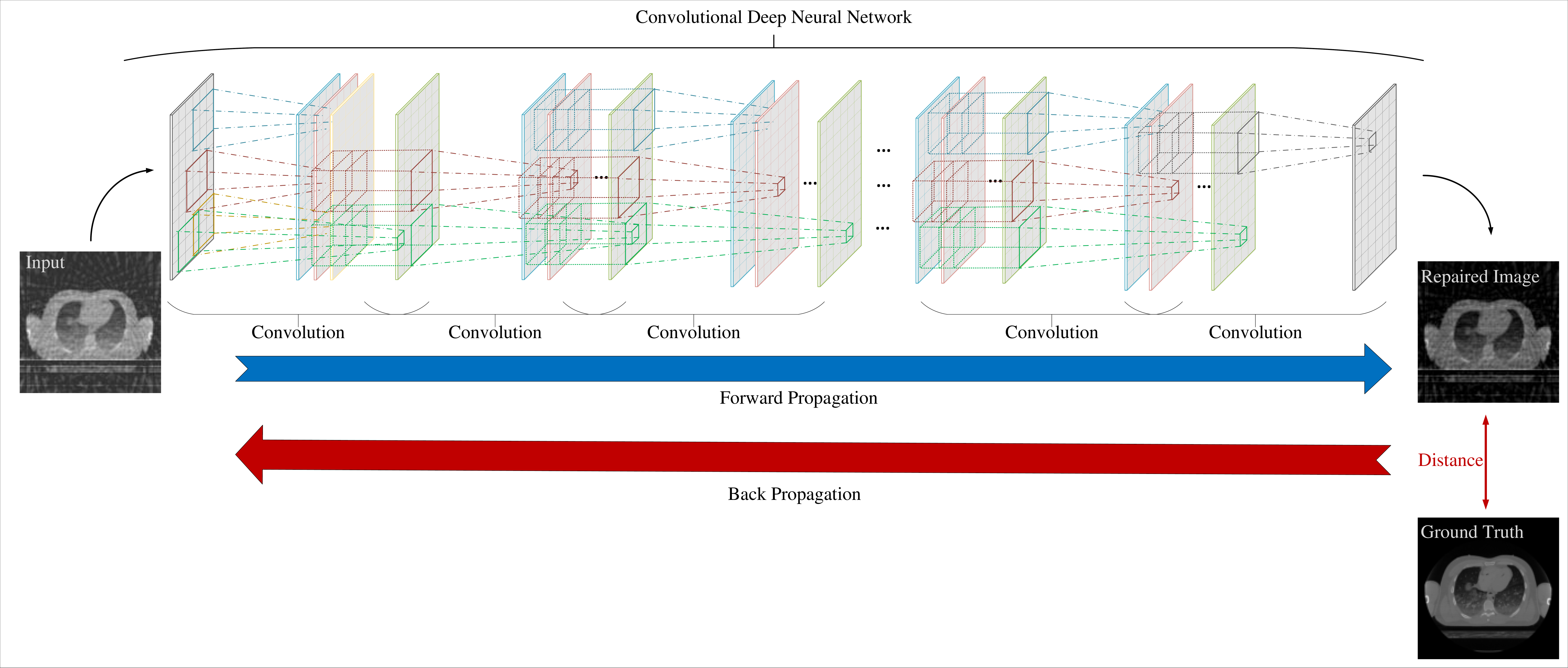}%
\caption{Fully Convolutional Deep Neural Networks accept and return images at their input and output nodes. These networks are popular at removing the reconstruction artefacts in low dose CT reconstructions.}
\label{fig:net}
\end{figure*}

Based on the observations and the discussions in the previous section, the marketing in the near future will not involve any end to end solution in the low dose CT reconstruction. In fact, the deep learning approach already found its way into the medical imaging industries. Regarding the article \cite{AICE}, the first industry level deep learning based low dose CT reconstruction method follows the first approach described in section \ref{sec:helpinghand}. In this method, a Convolutional Deep Neural Network is utilized to remove the artefacts from the conventional reconstructions.\\ 

Figure \ref{fig:net} shows a potential schematic of how the training is accomplished. The networks proposed in the literature often exploit fully convolutional architecture. A fully convolutional network is a neural network which just contains convolution, deconvolution, pooling, and unpooling layers. They usually take advantage of techniques such as Batch Normalization \cite{ioffe2015batch}, drop out \cite{dropout}, skipped connections and residual blocks \cite{RESNET}. These methods are employed to improve training convergence, avoid overfitting and keep high-frequency information from the input signal. After each convolutional operation the signal passes through an activation function also knows as nonlinearity. The most favorite nonlinearity is the ReLU \cite{RELU} and its variations such as leaky ReLU \cite{LEAKYRELU}, and ELU \cite{ELU}.\\
  
The training is accomplished by two main steps. Forward propagation and backpropagation. In the forward propagation step, the low-quality image is passed through the network and the output is aquired by calculating a set of convolutions. The next step is to compare the output of the network with its corresponding ground truth.  Loss value is calculated using one or more distance functions, designed based on the problem characteristics. In the method shown in figure \ref{fig:net} the most popular distance measure used in the literature is the mean squared error given by:

\begin{equation}
Loss = \frac{1}{B_s H W}\sum_{k=1}^{B_s}\sum_{j=1}^{H}\sum_{i=1}^{W}\big(O(i,j,k)-t(i,j,k)\big)^2.
\label{eq:MSE}
\end{equation}

Wherein $W$, $H$, and $B_s$ are width, height and the batch size of the input signal, $O$ and $t$ are the output signal and the target values respectively.\\

The next step is known as backpropagation. In this step, the derivative of the loss function with respect to every network parameter is calculated and these values are updated using the calculated gradient to decrease the loss.  
There are several different methods in performing the update which are mostly based on gradient descent technique such as Stochastic Gradient Descent (SGD), SGD with Nestrov Momentum \cite{SGDNEST}, AdaGrad \cite{ADAGRAD}, RMSprop \cite{RMSPROP}, and ADAM \cite{ADAM}. Although the most popular optimization method is ADAM since it assigns a learning rate and a momentum for each learnable parameter. The update step of ADAM is as follows:

\begin{equation}
\theta_n = \theta_{n-1}-\frac{\eta}{\sqrt{\hat{\nu}+\epsilon}}\hat{m}_n
\end{equation}

where

\begin{equation}
\hat{m}_n=\frac{m_n}{1-\beta_1}~~~~~,~~~~~\hat{\nu}=\frac{\nu_n}{1-\beta_2}
\end{equation}

and

\begin{equation}
m_n = \beta_1m_{n-1}+(1-\beta_1)\nabla_\theta F(\theta) 
\end{equation}
\begin{equation}
\nu_n = \beta_2\nu_{n-1}+(1-\beta_2)\nabla^2_\theta F(\theta)
\end{equation}

where and $\theta_n$ is the set of parameters in iteration $n$, $F$ is the model’s mapping function and, $\eta$ is the learning rate. The default values for $\beta_1$, $\beta_2$ and, $\epsilon$ are 0.9, 0.999 and $10^{-8}$, respectively.
This optimization method presents a solution to problems such as vanishing learning rate, slow convergence, and fluctuations in the cost value.\\

A large community of programmers and machine learning engineers are dedicated to research and develop faster and more versatile software platforms for Deep Learning use cases. There are many examples such as Tensorflow \cite{TENSORFLOW}, Theano \cite{THEANO}, Pytorch \cite{PYTORCH}, and MXNET \cite{MXNET} which provide an appealing experience and helpful interfaces to train and test DNNs with fast and memory efficient implementations. Almost every framework include Convolution, Deconvolution, Pooling , Fully Connected, Batch Normalization and Drop out techniques with almost all popular optimization methods implemented. In the near future the medical imaging industry in general and CT reconstruction in particular moves to a new level with the approval of Machine Learning techniques and their amazing outcomes.

\section{Conclusion}
In this article, the low dose CT reconstruction problem has been investigated from a machine learning point of view. The CT imaging, reconstruction methods, low dose problem and advantage and disadvantage of model-based approaches were covered as well.\\

The deep neural networks play an important role in our modern world technology development and like every other signal processing branches they vastly influence medical imaging science. Currently, the academic and the industrial worlds point at the DNNs as a solution for the low dose CT imaging. The main reason is the lack of a single representative model for such a problem. Deep Learning provides powerful tools in learning the artefacts caused by low SNR or incomplete set of observations in the sinogram signal. Two main approaches have been proposed in the literature in using data-driven methods to perform the reconstruction. The first one is to exploit these techniques to remove the noise and artefacts from the reconstructed image and the second one is to train an end to end model to translate sensor signal data straight into the image space representation. Both methods were investigated in the current article and concluded that the first approach gives more satisfactory results in the low dose CT use cases for two reasons:

\begin{enumerate}
\item At the end to end approach, the size of the model grows exponentially with respect to the size of the input signal which increases the chance of overfitting and also the implementation becomes nontrivial in real life scenarios with the requirements for high-resolution acquisitions.\\ 

\item The end to end model totally ignores the model geometry. This raises into larger problems in low dose cases where the underlying problems are highly ill-posed. Based on the observations concluded in section \ref{sec:DNNinCT}, this information plays a crucial role in retaining details in the reconstructed image.\\
\end{enumerate}

Industrial marketing is in favor of products from the first approach with using DNNs as helping hands for model-based methods. And also from the visual observations of the results presented in \cite{AICE}, these products compensate for lowering the X-ray tube current rather than limited or incomplete projection scenarios.
The future for these technologies is bright with affordable parallel processing hardware and open source and free software and they play an important role in developing a more accurate and higher quality medical acquisitions.


%



\section*{Acknowledgment}

This work is financially supported by VLAIO (Flemish
Agency for Innovation and Entrepreneurship), through the
ANNTOM project HBC.2017.0595.

\ifCLASSOPTIONcaptionsoff
  \newpage
\fi



%

\bibliographystyle{IEEEtran}
\bibliography{bibFinal}



%

\begin{IEEEbiographynophoto}{Shabab Bazrafkan}
received his B.Sc degree from Urmia University, Urmia, Iran in electrical engineering in 2011, M.Sc degree from Shiraz University of Technology (SuTECH) in telecommunication engineering, Image processing branch in 2013 and Ph.D. from the National University of Ireland, Galway (NUIG) in Deep Learning and Neural Network design in 2018 and he is currently a postdoctoral researcher working on low dose CT image reconstruction using machine learning techniques with VisionLab at the University of Antwerp.
\end{IEEEbiographynophoto}

\begin{IEEEbiographynophoto}{Vincent Van Nieuwenhove}
received his master’s degree in Physics in 2013 at the University of Antwerp, Belgium with a thesis on statistical processing of functional MRI data. Afterwards, he pursued his PhD at the imec-Vision lab, University of Antwerp. He received his PhD in Physics in 2017 with a thesis entitled ‘Model-based reconstruction algorithms for dynamic X-ray CT’. In 2018, Vincent joined Agfa NV, Belgium as ‘Research Engineer in 2-3D Medical Imaging’.

\end{IEEEbiographynophoto}

\begin{IEEEbiographynophoto}{Joris Soons}
received his M.Sc degree in physics from University of Antwerp, Belgium in 2007. During his PhD (2007-2012) at the lab of biomedical physics (University of Antwerp) and as postdoctoral researcher (2012-2015) at the Otobiomechanics group (Stanford University, USA), he focussed on 3D imaging techniques, modelling and mechanical experiments in biomechanics. Currently he is a researcher in 3D reconstruction and image processing at AGFA NV, Belgium.
\end{IEEEbiographynophoto}

\begin{IEEEbiographynophoto}{Jan De Beenhouwer}
obtained a M.Sc. in Computer Science Engineering in 2003 from the KU Leuven,
Belgium and a Ph.D. in Biomedical Engineering from the University of Ghent, Belgium in 2008. He was a postdoctoral fellow for 2 years at the same institution prior to joining the Vision Lab at the University of Antwerp, Belgium. From 2012-2016 he was a research manager for Vision Lab at the strategic research centers iMinds and IMEC. Since 10/2018 he is appointed as a research professor
and leads the ASTRA group in imec-Vision Lab which focuses on the development of advanced
computational methods for tomography as well as new reconstruction techniques that lead to better
reconstruction quality compared to classical reconstruction methods. His main interest is in image
reconstruction, processing and analysis with focus on computed tomography and electron
tomography.
\end{IEEEbiographynophoto}

\begin{IEEEbiographynophoto}{Jan Sijbers}
graduated in Physics in 1993. In 1998, he received a PhD in Physics from the University of Antwerp, entitled Signal and Noise Estimation from Magnetic Resonance Images". He was an FWO Postdoc at the University of Antwerp and the Delft University of Technology from 2002-2008. In 2010, he was appointed as a senior lecturer at the University of Antwerp. In 2014, he became a full professor. He is the head of imec-Vision Lab, a research lab focusing on image reconstruction, processing, and analysis. His main interests are in the domain of Magnetic Resonance Imaging and X-ray Computed Tomography. He is Senior Area Editor of IEEE Transactions on Image Processing as well as Associated Editor of IEEE Transactions on Medical Imaging.
\end{IEEEbiographynophoto}




\end{document}